\documentclass[12pt]{spieman}  
\usepackage{graphicx}
\usepackage{subfigure} 
\usepackage{authblk}
\usepackage{mathptmx}
\usepackage{amsmath}
\usepackage{xcolor}
\usepackage{graphicx}
\usepackage{multirow}
\usepackage{listings}
\usepackage{multicol}
\usepackage{minted}
\usepackage{tabularx}
\usepackage{makecell}

\definecolor{codegreen}{rgb}{0,0.6,0}
\definecolor{codegray}{rgb}{0.5,0.5,0.5}

\lstdefinestyle{mystyle}{
  numberstyle=\tiny\color{codegray},
  stringstyle=\color{codegreen},
  keywordstyle=\color{black}\bfseries,
  breakatwhitespace=false,         
  breaklines=true,                 
  keepspaces=true,                 
  showspaces=false,                
  showstringspaces=false,
  showtabs=false,                  
  tabsize=2
}
\usepackage{times}
\usepackage{parskip}
\usepackage{lineno}

\usepackage{array}
\newcolumntype{P}[1]{>{\centering\arraybackslash}p{#1}}
\usepackage{booktabs} 
\usepackage{diagbox}

\title{Accelerating mesh-based Monte Carlo simulations using contemporary graphics ray-tracing hardware}

\author[1]{Shijie Yan}
\author[1]{Douglas Dwyer}
\author[1]{David R. Kaeli}
\author[4,1,*]{Qianqian Fang} 

\affil[1]{Northeastern University, Department of  Electrical and Computer Engineering, Boston, Massachusetts, United States.}
\affil[4]{Northeastern University, Department of Bioengineering, Boston, Massachusetts, United States.}

\affil[*]{Corresponding author: q.fang@neu.edu}

\begin{document} 
\begin{sloppypar}
\maketitle

\begin{abstract}

\textbf{Significance}:
Monte Carlo (MC) methods are the gold-standard for modeling light-tissue interactions due to their accuracy. Mesh-based MC (MMC) offers enhanced precision for complex tissue structures using tetrahedral mesh models. Despite significant speedups achieved on graphics processing units (GPUs), MMC performance remains hindered by the computational cost of frequent ray-boundary intersection tests.

\textbf{Aim}: 
We propose a highly accelerated MMC algorithm, RT-MMC, that leverages the hardware-accelerated ray traversal and intersection capabilities of ray-tracing cores (RT-cores) on modern GPUs. 

\textbf{Approach}: 
Implemented using NVIDIA's OptiX platform, RT-MMC extends graphics ray-tracing pipelines towards volumetric ray-tracing in turbid media, eliminating the need for challenging tetrahedral mesh generation while delivering significant speed improvements through hardware acceleration. It also intrinsically supports wide-field sources without complex mesh retesselation.

\textbf{Results}: 
RT-MMC demonstrates excellent agreement with traditional software-ray-tracing MMC algorithms while achieving 1.5$\times$ to 4.5$\times$ speedups across multiple GPU architectures. These performance gains significantly enhance the practicality of MMC for routine simulations.

\textbf{Conclusion}:
Migration from software- to hardware-based ray-tracing not only greatly simplifies MMC simulation workflows, but also results in significant speedups that are expected to increase further as ray-tracing hardware rapidly gains adoption. Adoption of graphics ray-tracing pipelines in quantitative MMC simulations enables leveraging of emerging hardware resources and benefits a wide range of biophotonics applications.
\end{abstract}


{\noindent \footnotesize\textbf{*}Address all correspondence to: Qianqian Fang, Email: \linkable{q.fang@neu.edu} }

\begin{spacing}{1}   

\section{Introduction} \label{rtmmc_intro}
The Monte Carlo (MC) method is widely regarded as the gold-standard for modeling light propagation in biological tissues because it provides accurate solutions to the radiative transfer equation (RTE) in a wide range of tissue types, including low-albedo media, where diffusion-based models often fail~\cite{Hielscher1998, Okada2003, Custo2006}. MC methods achieve this accuracy by simulating light-tissue interactions such as absorption, scattering, reflection, and transmission through the random sampling of their respective probability distributions~\cite{Wang1995}. However, the stochastic nature of MC simulations makes them computationally intensive, requiring large numbers of photon packets to attain stable solutions.

Since the release of the first open-source MC code, MCML~\cite{Wang1995}, which models tissues as multi-layered structures with one-dimensional (1D) variations in optical properties along the $z$-axis, substantial progress has been made in utilizing more physiologically realistic tissue models in MC simulations. These advances often involve refining tissue representations using more sophisticated geometries. For example, voxel-based MC (VMC) methods account for three-dimensional (3D) heterogeneity by discretizing tissue domain into cubic voxels, each representing a specific tissue type~\cite{Boas2002, Fang2009mcx, mcxyz, Marti2021}. However, voxelated representations of surfaces often struggle to conform to curved boundaries, leading to significant simulation errors~\cite{Binzoni2008, TranYan2020, Yan2020}.

To improve the accuracy of the boundary, various hybrid approaches have been proposed. Enhanced voxel-based methods integrate additional geometric details, such as surface normals~\cite{Tran2020} or sub-voxel surfaces~\cite{Yan2020}, to improve boundary representation. The method proposed in Tran and Jacques~\cite{Tran2020} enhances the modeling of reflection and refraction at curved boundaries but does not resolve the partial volume effects caused by voxel rasterization~\cite{Yan2020}. In contrast, the approach proposed by Yan and Fang~\cite{Yan2020} introduces an implicit oblique surface within each boundary voxel to capture sub-voxel heterogeneity, effectively mitigating partial volume effects as well as errors in reflection and refraction. However, this technique still fails to capture microscopic structures that are critical for multi-scale imaging~\cite{Yuan2020}. Other strategies, such as incorporating signed distance functions (SDF) into boundary elements~\cite{McMillan2022}, have also been explored to improve boundary accuracy. Alternatively, shape-based MC methods define 3D objects using parametric functions, offering high accuracy for simple geometries such as spheres, cylinders, and cuboids~\cite{Majaron2015, Ding2016, Periyasamy2014, Zhang2015}. While these methods achieve precise boundary modeling, they lack the flexibility to accommodate arbitrarily complex geometries.

Given these limitations, triangular and tetrahedral mesh-based MC (MMC) methods have emerged~\cite{Margallo2007,Fang2010}, drawing parallels between photon MC simulations and rapidly advancing fields such as computer 3D graphics~\cite{Fang2012}. Ray-tracing and ray-casting problems lie at the core of both MC photon simulations and modern graphics rendering techniques. To render a 3D scene, modern graphics rendering engines cast rays from the center of each display pixel and perform ray-tracing to bounce the rays through the 3D scene to compute the color (red, green, blue, and alpha/transparency) to be rendered at each pixel. To achieve fast rendering speeds, highly simplified physics is often applied to graphics-oriented ray-tracing computations, with rays interacting primarily with surface-associated optical properties through physics-based rendering using bidirectional scattering distribution functions (BSDFs)~\cite{pharr2023physically}. The handling of volumetric ray-casting, commonly known as subsurface scattering (SSS)~\cite{jensen2001practical}, has been largely avoided due to its high computational cost and has only become viable in recent years.

In comparison, a biophotonics MC simulation is by nature a 3D volumetric ray-tracer that rigorously follows the physical laws of light-media interactions. It launches rays from the source and continuously performs ray generation-propagation-termination steps, following random sampling of the media's volumetric optical properties, including absorption coefficient ($\mu_a$ in mm$^{-1}$), scattering coefficient ($\mu_s$ in mm$^{-1}$), anisotropy ($g$) and refractive index ($n$). Instead of producing two-dimensional (2D) colored images as in most graphics ray-tracers, physics-based photon MC simulations output quantitative volumetric light distributions in the form of fluence (mm$^{-2}$) or fluence-rate (mm$^{-2}$s$^{-1}$).

The use of triangular meshes in representing complex scenes -- a common practice in graphics rendering -- has previously been extended to MC photon simulations~\cite{Margallo2007, Ren2010} to represent arbitrarily complex tissue and media boundaries. Triangular surfaces offer greater flexibility compared to planes (MCML~\cite{Wang1995}), voxels (VMC~\cite{Boas2002,Fang2009mcx}) or parametric surfaces (shape-based MC~\cite{Majaron2015} or SDF~\cite{McMillan2022}), allowing the tissue domain to be discretized into homogeneous regions. A notable overhead in surface-based MC photon propagation is that the ray-tracer must handle multiple ray-triangle intersection tests for every propagation step. To accelerate this process, triangle spatial partitioning schemes, such as kd-trees~\cite{bentley1975multidimensional}, octrees~\cite{meagher1982geometric}, or bounding volume hierarchies (BVH)~\cite{kay1986ray}, previously developed in the field of computer graphics, have also been adopted in MC simulations. Nevertheless, even with such acceleration structures (AS), advancing a ray (or a photon packet) by one step often requires testing over a dozen triangles, with the added overhead of traversing nested AS data structures in memory.

In 2010, two MC algorithms based on tetrahedral mesh models were published to specifically address this overhead of ray-triangle intersection tests~\cite{Fang2010, Shen2010}. In these approaches, a surface-mesh-based domain is further tessellated to partition the space enclosed between surfaces into tetrahedral elements, and the index of the tetrahedral element enclosing a photon's current position is continuously tracked at any given time in the simulation. As a result, only the 4 triangles of the bounding tetrahedral element need to be tested for every photon's movement, limiting the ray-triangle testing number per step to an average of 2.5~\cite{Fang2011}. A drawback of this approach is the requirement for tetrahedral mesh generation, which can be challenging for complex domains, such as human brains~\cite{TranYan2020}. Using MMC with a dense tetrahedral mesh could still be computationally expensive. Using coarsely tessellated tetrahedral meshes was explored in a dual-grid MMC (DMMC) algorithm~\cite{Yan2019dmmc} to significantly reduce the number of ray-tetrahedron intersection tests while simultaneously achieving improved output accuracy.

Over the past two decades, the wide availability of many-core processors, particularly graphics processing units (GPUs), has driven substantial performance improvements in MC simulations~\cite{Alerstam2008, Fang2009mcx, Ren2010, Doronin2011, Yu2018, Fang2019, Young-Schultz2019, Jonsson2020}. GPU-accelerated MC implementations have supported light simulations in complex media including multi-layered ~\cite{Alerstam2010}, voxelated~\cite{Fang2009mcx}, or tetrahedral meshes~\cite{Fang2019, Young-Schultz2019}. Despite these significant speed improvements using modern computing hardware, the desires to further improve MC simulation performance, both in speed and in media complexity, persist as the field of biophotonics expands into diverse applications.

Recent advances in real-time ray-tracing, largely driven by the computer graphics and gaming industries, result in the emergence of specialized ray-tracing (RT) hardware, known as the ``RT-cores''~\cite{OptiXGuide}, which has been widely supported by GPUs manufactured after 2020. Adapting MMC to utilize this emerging hardware~\cite{Kilgariff2018} offers new opportunities to further enhance the performance of quantitative 3D MC simulations. Inspired by these developments, here we present ``RT-MMC'', a highly accelerated MMC algorithm that integrates a contemporary graphics-based RT workflow. This approach capitalizes on modern RT hardware, offering excellent scalability as the performance of RT cores continues to improve rapidly. Although hardware-accelerated ray-tracing has been applied to particle transport problems in MC~\cite{Salmon2019}, its application has mainly been confined to traditional rendering tasks. To our knowledge, this work is the first to apply RT hardware in quantitative physics-based particle/photon transport simulations.

In this work, we implemented RT-MMC using NVIDIA OptiX -- a high-level compute unified device architecture (CUDA) based RT library designed for real-time rendering on modern NVIDIA GPUs~\cite{Kilgariff2018, Parker2010}. To optimize performance, we utilized OptiX's BVH acceleration structures (AS) to efficiently partition triangular mesh based domain representations. By adopting OptiX's built-in AS and triangular primitives, RT-MMC fully leverages the hardware-accelerated ray traversal and ray-triangle intersection capabilities provided by the RT-cores. Although our implementation of RT-MMC is based on OptiX, our approach can be readily generalized to any ray-tracing framework, such as Microsoft DXR~\cite{DXRGuide}, Apple Metal~\cite{MetalRayTracing,VulkanRayTracingSIGGRAPH}, and Vulkan~\cite{VulkanRayTracingSIGGRAPH}.

In the remainder of this manuscript, we first give a brief introduction on the key components of the ray-tracing pipeline. Then we detail the implementation of RT-MMC, including different strategies in constructing AS, mapping photon simulation steps to ray-tracing events, facilitating widefield MMC simulations using ray-tracing pipelines, among other optimization strategies. We validate the proposed algorithm using the results of traditional MMC methods as references. Additionally, we compare performance across mesh domains of varying complexity and across different GPU architectures. Finally, we discuss our findings, the limitations of this work, and potential directions for future research.

\section{Background} \label{rtmmc_bg}
Modern ray-tracing (RT) pipelines require three key steps: 1) building AS, 2) writing ray propagation handling programs (known as ``shaders'') following each ray-tracing event, and 3) connecting the shader programs, ASes and other data resources to the ray-tracer engine using a specialized data structure known as shader binding table (SBT)~\cite{OptiXGuide, DXRGuide, VulkanGuide}.

\subsection{Acceleration structures}
In computer graphics, acceleration structures (AS) refer to data structures that help partition shape primitives (including triangles) in a scene to accelerate ray traversal and ray-triangle intersection testing. In previously published surface-based MMC algorithms~\cite{Margallo2007,Ren2010}, triangles are organized using manually created data structures to efficiently reject irrelevant geometry during ray traversal. For example, Margallo~\emph{et al.}~\cite{Margallo2007} applied a ``loose'' octree to recursively divide tissue volume into equal subvolumes. Similarly, Ren~\emph{et al.}~\cite{Ren2010} project triangle meshes and photon paths onto a 2D grid plane, restricting ray-triangle intersection checks to triangles within the grid cell intersected by the projected photon path. These methods rely on spatial subdivision to partition space and record objects within each defined region.

\begin{figure}[t]
\centering
    \subfigure[]{
        \includegraphics[width=0.30\textwidth]{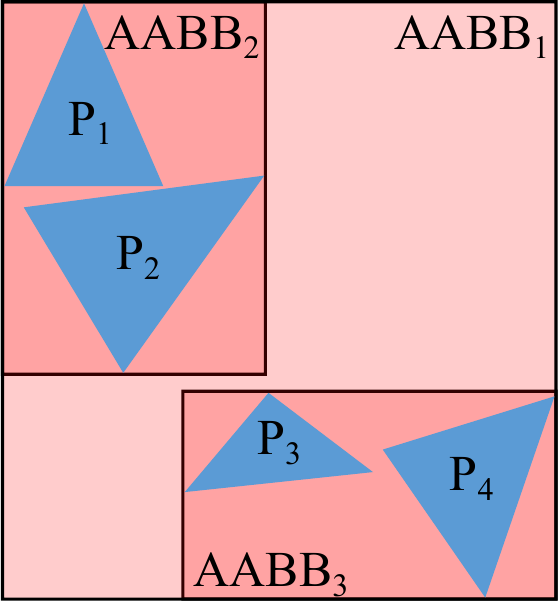}
        \label{rtmmc_bg_optix_fig_a}
    }
    \hfill
    \subfigure[]{
        \includegraphics[width=0.34\textwidth]{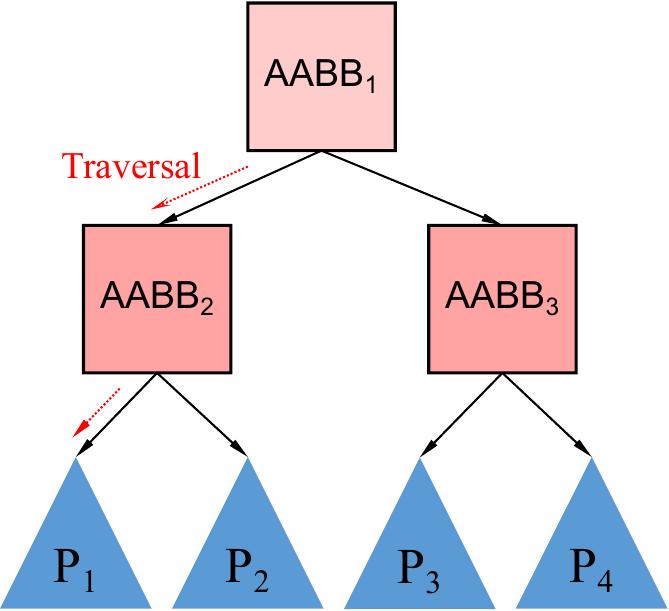}
        \label{rtmmc_bg_optix_fig_b}
    }
    \hfill
    \subfigure[]{
        \includegraphics[width=0.30\textwidth]{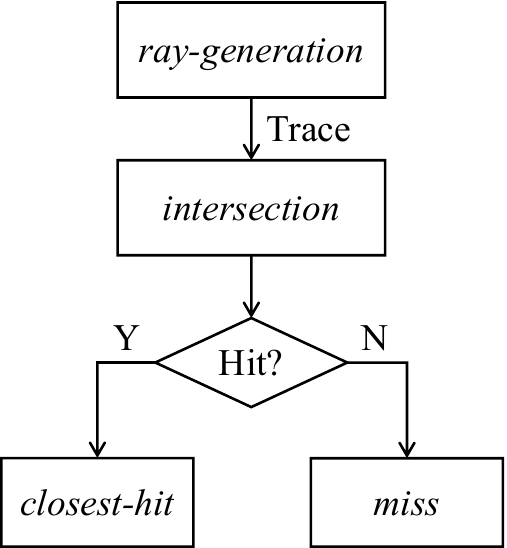}
        \label{rtmmc_bg_optix_fig_c}
    }
\caption{\label{rtmmc_bg_optix_fig} Ray-tracing acceleration structures (AS) and workflow, showing (a) spatial partition of shape primitives ($P_i$, $i= 1,\cdots,4$) using axis-aligned bounding boxes (AABBs) within a bounding volume hierarchy (BVH), (b) traversal of the AS, and (c) key ray-tracing event handling (shaders).}
\end{figure}

In contrast, modern RT pipelines employ BVH to organize geometries (see Fig.~\ref{rtmmc_bg_optix_fig_a}). Unlike spatial subdivision, BVH offers more flexibility by applying hierarchical grouping. It applies hierarchical grouping to all geometric objects and stores the bounding volume occupied by each group, thereby further accelerating geometry traversal. A more in-depth comparison of spatial and object subdivision methods can be found in~\cite{pharr2023physically}. The BVH is represented as a tree structure, where the internal nodes define bounding volumes that encompass their child nodes, while the leaf nodes contain the actual geometric primitives (see Fig.~\ref{rtmmc_bg_optix_fig_b}).

RT pipelines allow users to define the scene using triangular meshes, with the added capability of defining custom shape primitives (such as spheres and 3D curves~\cite{OptiXGuide}) alongside their corresponding axis-aligned bounding boxes (AABBs). Most ray-tracing software frameworks, including NVIDIA OptiX~\cite{OptiXGuide}, Microsoft DirectX Raytracing (DXR)~\cite{DXRGuide} and Vulkan Ray Tracing~\cite{VulkanGuide}, provide application programming interfaces (APIs) as abstraction layers to construct and customize BVH using geometric data. On modern GPUs equipped with RT cores, contemporary RT pipelines automatically enable RT-core accelerated BVH traversal and ray-triangle intersection tests, offering a rare opportunity to capitalize on this powerful computing hardware in MC simulations.


\subsection{Ray-tracing workflow and programmable shaders}
RT pipelines provide generalized interfaces, known as shader programs, to define how rays interact with the domain. These shader programs handle key events [see Fig.~\ref{rtmmc_bg_optix_fig_c}] including ray-generation (``ray-gen'' shader), user-defined ray–primitive intersection (``intersection'' shader), detection of any intersection (``any-hit'' shader), closest-intersection (``closest-hit'' shader), and no-intersection (``miss'' shader). A ray here refers to any single straight segment along the tortuous path of the photon packet, defined by an origin and a direction. The ``ray-gen'' shader is responsible for iteratively casting the ray to the scene, retrieving and initializing the ray's states (positions, directions, weight, etc) between subsequent propagation segments. The intersection shader, if defined, allows users to write customized ray-primitive (spheres, curves, etc) intersection testing algorithms and report the distance of the intersection point from the ray's origin. When the shape primitives are triangles, an RT-capable GPU executes the built-in ray-triangle intersection shader on the RT-cores, offering hardware acceleration. Once all triangles within the AS are tested, the ``closest-hit'' shader is triggered, reporting the shortest distance among all intersections. If a ray fails to intersect any geometry, the ``miss'' shader is triggered to allow users to define fallback behaviors, such as rendering the background. The above steps are repeated for all photon propagation path segments, advancing the ray/photon packet from one position to the next, until it is terminated. These shaders provide fine-grained control over the ray-tracing process, allowing users to create complex lighting and material effects, including advanced physical property handling and simulations, without dealing with low-level GPU implementations.

\subsection{Shader binding table}
The shader binding table (SBT) is a crucial component in the RT pipeline, providing the mechanism to link geometric primitives with their associated shaders~\cite{Marrs2021}. It binds ray-tracing shader programs to each geometry within the AS~\cite{VulkanGuide} and includes additional resources such as material properties, textures, or other data that the shaders access during execution. During ray-tracing, the SBT is referenced whenever a ray interacts with geometry or misses objects. This ensures that the appropriate shader is invoked for each event. By decoupling the computational logic from the scene geometry, the SBT provides an efficient abstraction for event-based handling of complex and diverse ray-tracing simulations while maintaining high performance.

\section{Methods} \label{rtmmc_methods}
\subsection{Domain and mesh data preparation -- building AS and SBT} \label{rtmmc_methods_as}
\begin{figure}[t]
\centering
    \subfigure[]{
        \includegraphics[width=0.30\textwidth]{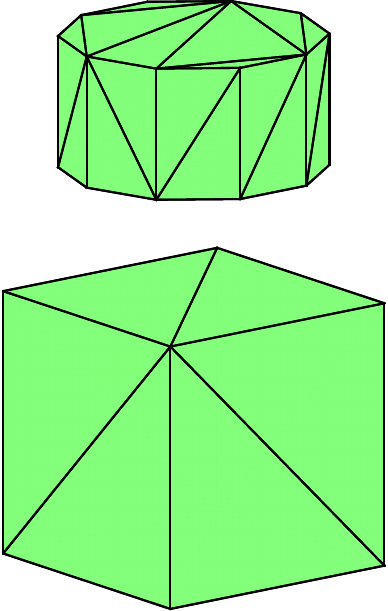}
        \label{rtmmc_methods_as_fig_a}
    }
    \hfill
    \subfigure[]{
        \includegraphics[width=0.30\textwidth]{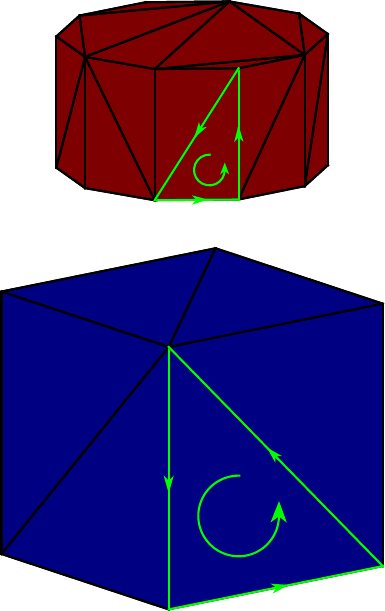}
        \label{rtmmc_methods_as_fig_b}
    }
    \hfill
    \subfigure[]{
        \includegraphics[width=0.30\textwidth]{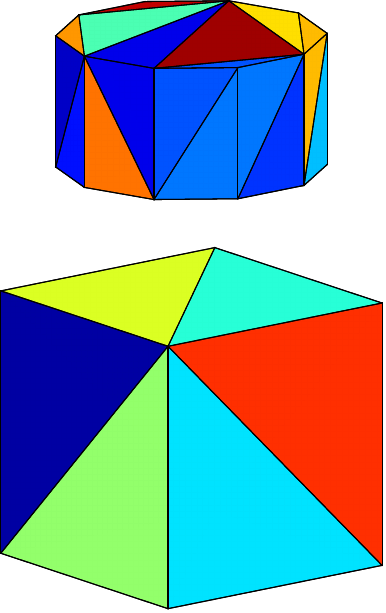}
        \label{rtmmc_methods_as_fig_c}
    }
\caption{\label{rtmmc_methods_as_fig} Illustrations of three mesh partitioning strategies for constructing acceleration structures (AS): (a) Single-AS, (b) Region-AS, and (c) Tetra-AS, for a scene containing two disconnected geometries -- a cylinder and a cubic object. We label each AS with a unique color: in (a), a single AS is used; in (b), two ASes are labeled; in (c), the interior space of each object is further tessellated into tetrahedral elements, and one AS is created per tetrahedron. Green arrows denote the winding order of the triangle vertices that defines its front-face.}
\end{figure}

In this study, we investigate three different strategies for constructing the AS. The first approach, referred to as ``Single-AS'', involves creating a single AS that encompasses all triangles. The second approach, referred to as ``Region-AS'', constructs multiple ASes, each containing triangles that form a ``simply connected surface''~\cite{botsch2010polygon}. The third approach, referred to as ``Tetra-AS'', involves tessellating the enclosed space into tetrahedral elements and defining individual AS for each element to mimic the behavior in traditional MMC algorithms~\cite{Fang2010,Shen2010}. These strategies are depicted in Fig.~\ref{rtmmc_methods_as_fig}, and their key characteristics are summarized in Table~\ref{rtmmc_methods_summary_table}.

In the Single-AS approach [Fig.~\ref{rtmmc_methods_as_fig_a}], the entire mesh, including all triangular surfaces at all tissue boundaries, is used to create a single AS. As noted earlier, the construction of the AS is managed by the GPU hardware. In comparison, the Region-AS approach [Fig.~\ref{rtmmc_methods_as_fig_b}] is designed to provide refined control over the spatial division of the BVH~\cite{Stich2009}. In this method, the triangular surfaces of each tissue region enclosed by a simply connected surface are grouped to form an individual AS. Ray-tracing within a region only requires testing triangles of the region-specific AS, potentially reducing the number of triangles to be tested. The Tetra-AS approach (Fig.~\ref{rtmmc_methods_as_fig_c}) adopts a partitioning strategy inspired by traditional tetrahedral MMC algorithms~\cite{Fang2010, Shen2010}. In this method, we tessellate the domain into tetrahedral elements, with a separate AS per tetrahedron. This fine-grained partition strategy is aimed at limiting the maximum number of ray-triangle testing to 4, closely mimicking that in the ray-tracing calculations in a tetrahedral-mesh based MMC algorithm~\cite{Fang2011,Fang2010}. For both Region-AS and Tetra-AS, triangle vertices within each connected region or tetrahedral element are arranged with a counter-clockwise winding~\cite{OptiXGuide} to ensure that their front-faces point outward, as indicated by the green arrows in Fig.~\ref{rtmmc_methods_as_fig_b}. However, a notable trade-off in these two approaches is the increased memory usage compared to Single-AS, as triangles shared between adjacent regions or tetrahedra are duplicated across the corresponding AS instances.

Across all three AS construction strategies, the SBT provides additional boundary information for handling light-boundary behaviors and tracking tissue types. For all strategies, the SBT includes precomputed surface normal vectors, which are essential for modeling light reflection and transmission. In the Single-AS approach, the winding order of each triangle is not specified. In addition to the surface normals, the SBT stores the tissue types associated with the front and back faces (defined by the winding order of the vertices~\cite{OptiXGuide}), allowing convenient updates of medium information when a photon crosses a boundary, depending on which face is intersected. In contrast, for the Region-AS and Tetra-AS approaches, since photons only intersect the inward-facing surfaces of the enclosing tissue domain and all front faces are ensured to be outward-facing during AS construction, we apply front-face culling~\cite{OptiXGuide} to further reduce the number of ray-triangle tests. In these two approaches, the SBT also provides the traversable AS handle and the tissue type of the adjacent AS instance when a photon crosses a boundary. The key differences in SBT construction across the three RT-MMC algorithms are summarized in Table~\ref{rtmmc_methods_summary_table}.

\subsection{Implementing photon migration using ray-tracing pipelines} \label{rtmmc_methods_programs}
\begin{figure}
\centering
{\includegraphics[width=\textwidth]{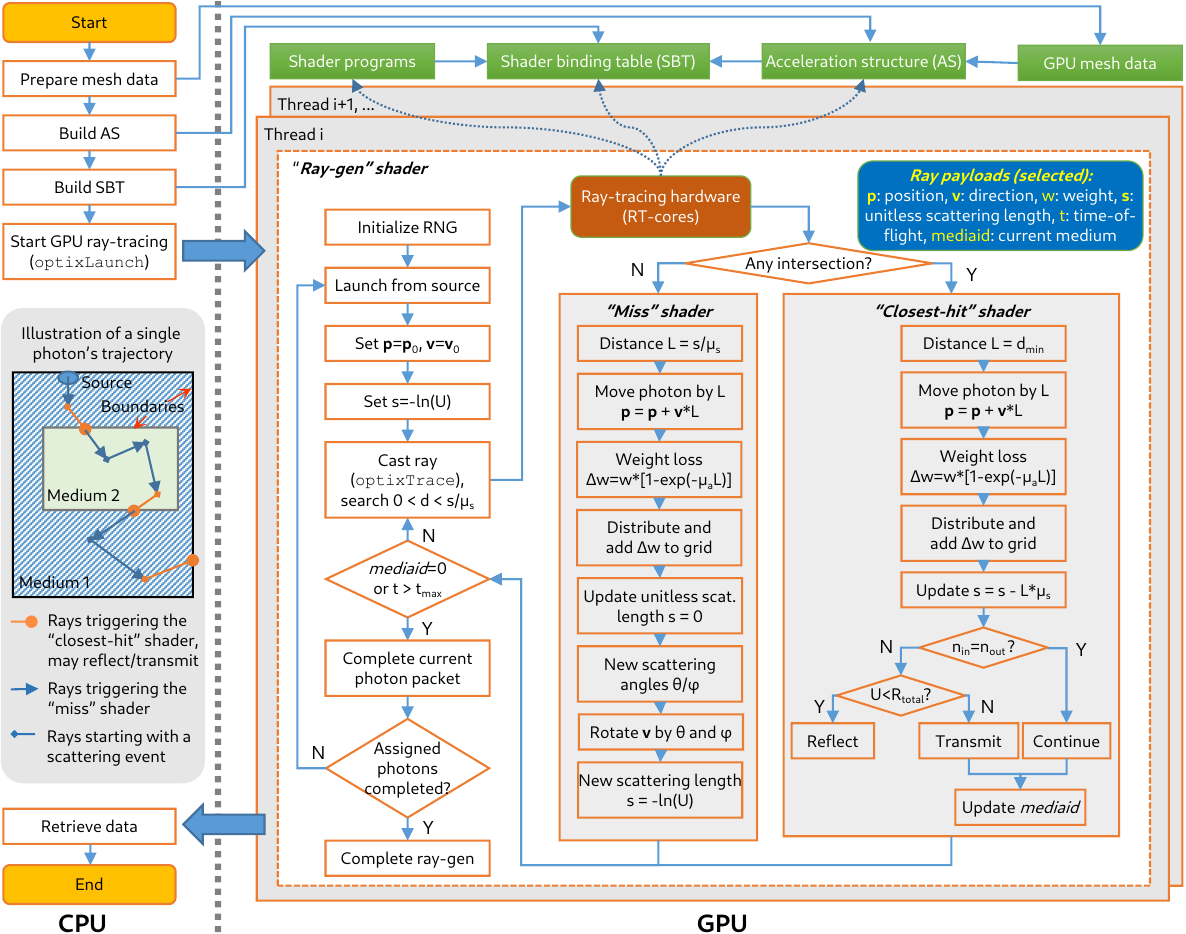}}
\caption
{\label{rtmmc_methods_programs_fig}
Diagram describing the workflow of a ray-tracing hardware accelerated Monte Carlo (MC) photon simulation. An illustration of the trajectory of a single photon packet is shown in an inset. A parallel xorshift128+ random number generator (RNG) is implemented in each thread; $U$ denotes a uniformly distributed random number between 0 and 1.}
\end{figure}

In Fig.~\ref{rtmmc_methods_programs_fig}, we present a massively-parallel MC light transport algorithm implemented using the contemporary RT pipeline. The algorithm starts on the central processing units (CPU) -- first processing and uploading mesh data to the GPU's global memory, building AS and SBT, then calling the OptiX's \texttt{optixLaunch} API to create a 1D array of threads on the GPU, with each thread invoking a ray-gen shader. The number of threads is determined by the hardware's capabilities, using heuristics similar to those used in our previous GPU-accelerated MC algorithms~\cite{Yu2018, Fang2019}. Each thread is also assigned a per-thread photon number by evenly dividing the total photon number by the number of threads. In RT-MMC, a parallel xorshift128+~\cite{Vigna2017} random number generator (RNG) is implemented to provide fast and long-period ($2^{128}-1$) random numbers for the MC simulation in each thread.

\subsubsection{Photon state management -- ray payloads}

The NVIDIA OptiX based RT pipeline allows users to define up to 32$\times$ 4-byte integers/floating point variables per ray~\cite{OptiXGuide} to track and update its state (such as position, direction, etc). In the ray-tracing literature, these are known as ``ray payloads''. Ray payloads are directly stored in the register space -- the fastest memory in the GPU~\cite{OptiXGuide}. These payloads can be conveniently passed between different shader programs to allow them to communicate and update the ray information.

In the Single-AS approach, 14 payloads are allocated to represent photon states. These include positions ($\mathbf{p}=\{x,y,z\}$, 3$\times$ payloads), directional cosines ($\mathbf{v}=\{v_x,v_y,v_z\}$, 3$\times$ payloads), the remaining unit-less scattering length ($s$, 1$\times$ payload), packet weight ($w$, 1$\times$ payload), time-of-flight ($t$, 1$\times$ payload), the enclosing medium type/label ($mediaid$, 1$\times$ payload), and RNG states (4$\times$ payloads). For the Region-AS and Tetra-AS approaches, two additional payloads are required to track the 64-bit traversable handle of the AS enclosing the ray's origin (see Table~\ref{rtmmc_methods_summary_table}).

\subsubsection{Photon propagation management -- ray-gen shader}

In an MC simulation, a photon's propagation typically follows a tortuous path because of the presence of scattering. All scattering sites divide the photon paths into numerous directional line segments, each connecting between the adjacent scattering sites. These linear path segments are further truncated into even finer segments delineated by discretized tissue boundaries, including layered, voxelated, surface, or tetrahedral discretization elements. The full path of a photon's trajectory is referred to as a ``primary ray''; the finest straight-line segment is referred to as a ``secondary ray'' or a ``ray''. A picture showing the subdivisions of the photon path is shown as an inset in Fig.~\ref{rtmmc_methods_programs_fig}.

In our RT-MMC implementation, we use the ``ray-gen'' shader to simulate the entire path (i.e. primary ray) of a photon packet's propagation. Every photon packet is launched at the source by initializing its launch position ($\mathbf{p}=\mathbf{p_0}$) and direction ($\mathbf{v}=\mathbf{v_0}$) according to the specified source. The initial launch also sets the unit-less scattering distance $s$ based on the exponential distribution, i.e. $s=-\textrm{log}(U)$, where $U$ is a uniformly distributed random number between 0 and 1. After the initial launch, the packet iteratively traverses the discretized domain, advancing from one scattering site to the next. Finally, the packet is terminated when it exits the domain or exceeds the maximum time limit ($t_{max}$). Compared to our CUDA and OpenCL based MC simulation workflows, the ``ray-gen'' shader is similar to a GPU kernel and runs in a single GPU thread. A diagram showing the detailed steps in the ray-gen shader can be found in Fig.~\ref{rtmmc_methods_programs_fig}.

Because of the randomness of the photon propagation in complex media, the length and the corresponding simulation time for each photon packet may vary dramatically. To optimize the workload balance of each GPU thread, our ray-gen shader simulates a batch of photon packets sequentially. The number of photo packets simulated in each ray-gen shader is determined by dividing the total simulated packet number by the number of threads. This strategy was also used in our CUDA~\cite{Fang2009mcx} and OpenCL~\cite{Yu2018} based MC simulators.

\subsubsection{Photon scattering and absorption handling -- miss shader}

In RT-MMC, we handle photon's scattering and absorption events in the ``miss'' shader (see the middle part of Fig.~\ref{rtmmc_methods_programs_fig}). This is because for each ray (i.e. a photon path segment) we cast to the scene, we set a maximum ray length $d_{max}$ determined by the remaining scattering length using $d_{max}=s/\mu_s$, where $\mu_s$ (mm$^{-1}$) is the scattering coefficient of the medium. Assuming the ray stays within the simulated domain (see the exception in Section~\ref{rtmmc_results_future}), when no intersecting triangle is reported by the ray-tracing engine (thus, the miss shader is invoked), it means that the photon packet has arrived at its next scattering site before encountering tissue boundaries. In this case, we can update the photon's position based on the remaining photon scattering length. 

After completion of the current scattering path, a new scattering event is triggered. For each scattering event, 3 random variables are calculated: 1) the scattering zenith angle $\theta$ using the Henyey-Greenstein phase function~\cite{Wang1995}, 2) the azimuthal angle $\varphi=2\pi\times U$, and 3) a new value of $s$ is given by $s=-\textrm{log}(U)$. The computed new direction and scattering length are immediately updated in the ray's payloads.

As the photon's position has advanced, we should also consider the photon weight's attenuation due to the absorption of the current medium. The energy-loss due to absorption are then accumulated over a dual-grid voxelated data structure as previously reported~\cite{Yan2019dmmc}, which ultimately forms the volumetric fluence output of the MC simulation.

\subsubsection{Photon-tissue boundary intersection and absorption handling -- closest-hit shader}

For every ray casted to the domain, represented by one or multiple ASes, the RT-core automatically traverses the AS and identifies candidate triangles that may possibly intersect with the ray within the specified maximum scattering length ($d_{max}$), and runs ray-triangle intersection testing using dedicated RT-core hardware. For the $i$-th triangle that actually intersects the ray, the distance $d_i$ from the ray's origin to the intersection point is reported. After testing all candidate triangles, the ray-tracing engine compares all reported distances to the intersection points, invokes the ``closest-hit'' shader (see Fig.~\ref{rtmmc_methods_programs_fig} right side) and returns 1) the shortest intersection distance $d_{min} = \mathrm{min}(\{d_i\})$, and 2) the handles to the AS and the triangle that the ray encounters first, and 3) whether the front (see Fig.~\ref{rtmmc_methods_as_fig_b}) or the back face of the triangle is hit by the ray. As a result, we implement all photon-tissue boundary handling computations in the closest-hit shader program.

An important photon-tissue interaction is the handling of reflection and transmission at the tissue boundary. Using the intersecting triangle's index returned by the closest-hit shader, we can retrieve a pre-computed data structure, stored in a global memory buffer linked in the SBT (see the precomputed data for the SBT in Section~\ref{rtmmc_methods_as}), containing the normal vector of the triangle, as well as the medium label of the tissue adjacent to the intersecting triangle. Using the retrieved medium labels, we can determine whether the refractive indices ($n$) across the triangle have a mismatch. When a mismatch is detected, we can then use the current direction of the ray and the normal vector of the triangle to compute the incident angle and the total reflection coefficient ($R_{total}$). A 0-1 uniformly distributed random number $U$ is computed and compared to $R_{total}$. If $U < R_{total}$, the photon packet takes the reflection path; otherwise, it takes the transmission path. In either case, the photon's directional vector is updated according to Fresnel's law. When a photon takes the transmission path or hits a $n$-matched boundary, the ray's enclosing medium label is updated using the medium information associated with the intersecting triangle. For the Region- and Tetra-AS approaches, since the tissue domain is partitioned into multiple ASes, the AS handle must also be updated along with the medium ID to enable intersection tests within the next AS (see Table~\ref{rtmmc_methods_summary_table}). 

Regardless of whether a mismatch of $n$ is detected, the current position of the photon packet advances to the intersection point based on the closest-hit distance ($d_{min}$) returned. Because a photon package moves forward by the above distance, we again compute the energy loss of the packet weight using the current medium's absorption coefficient and accumulate the lost energy to a dual-grid voxelated memory buffer.

\subsubsection{Photon termination and re-launch -- ray-gen shader}

After executing the closest-hit or miss shaders, the updated photon states are returned to the ray-gen shader. The ray-gen shader continues to iteratively cast rays into the relevant ASes in the scene, propagating the photon as long as it remains ``alive''. A photon packet is considered ``dead'' and terminated once it either exits the tissue domain or reaches the maximum time-of-flight ($t > t_{max}$). Once a photon is terminated, the ray-gen shader launches the next photon packet from the source until the assigned maximum number of photons to the thread has been completed.

\subsubsection{Simulating wide-field and volumetric sources in a mesh-based domain}

Simulating a wide-field source is particularly challenging in traditional MMC. This is because traditional MMC requires one to know the tetrahedral element that encloses the photon at every step of propagation. Consequently, the tetrahedral element that contains the launch position of the photon must also be determined at the start. For point sources such as pencil beams, the launch position is fixed, and the index of the enclosing tetrahedron can be precomputed. In contrast, wide-field sources compute the initial position of each photon at runtime, introducing additional overhead to locate the initial enclosing element.

The problem becomes even more difficult when a wide-field source lies outside or partially outside the outer surface of the mesh-based simulation domain. To ensure that every photon is enclosed by a tetrahedron, traditional MMC requires an one-time preprocessing step -- mesh retessellation~\cite{Yao2016} -- to tessellate the space between the external 3D source aperture and the target mesh surface. Although we have developed automated mesh-generation tools~\cite{Yao2016} to streamline this retessellation process for wide-field MMC simulations, the mesh preprocessing overhead and the lack of robustness when tessellating complex geometries significantly limit the practicality and general applicability of this approach.

Adopting RT pipelines for MMC modeling greatly simplifies the simulation of wide-field sources. First, for external or partially external sources, RT-based MMC completely eliminates the need for mesh retessellation. This is because RT pipelines inherently support ray tracing from arbitrary starting positions. With hierarchical geometry organization using AS and hardware-accelerated AS traversal, RT pipelines provide fast and robust ray–triangle intersection testing regardless of the relative placement of the source and the domain.

Second, RT pipelines also greatly simplify the simulation of spatially distributed internal sources, including 3D volumetric sources inside a mesh domain. In conventional MMC, such sources require manually labeling all tetrahedra that encompass the distributed source, as well as scanning all candidate tetrahedra for every photon launched. Using a combination of AS domain representation and hardware based ray-triangle testing, RT pipelines can launch photons from spatially distributed sources from any location inside the domain without any additional preprocessing.

In this work, We extended the Single-AS RT-MMC to support photons launched from arbitrary locations, regardless of their placement relative to the tissue domain, to simulate wide-field sources. In the modified ``ray-gen'' shader, when a photon is launched without a pre-defined medium type, we cast an auxiliary ray into the scene with $d_{max}$ set to infinity (single-precision float max). If a ``miss'' is reported, the ray does not intersect the tissue domain (with exceptions noted in Section~\ref{rtmmc_results_future}), and the photon is terminated immediately. If an intersection is detected, the ``closest-hit'' shader retrieves the initial medium type from the associated SBT entry, using front- or back-face information depending on which side of the triangle is intersected. Once the initial medium type is determined, the ``ray-gen'' shader proceeds with standard RT-MMC propagation.

This modification eliminates the need to precompute the initial medium type for each photon and instead determines it dynamically at launch time using RT-core–accelerated ray traversal and intersection testing. Moreover, because the initial medium type is no longer determined by the enclosing tetrahedral element, this method avoids the need for mesh retessellation to force the mesh to enclose the source~\cite{Yao2016}. We implement this enhancement specifically in the Single-AS RT-MMC because it provides the most straightforward workflow: the entire tissue domain resides within a single AS, removing the need to determine an initial AS before casting the query ray.

\begin{table}[t]
\renewcommand*{\arraystretch}{1.5}
\begin{center}
\renewcommand{\tabularxcolumn}[1]{m{#1}}
\begin{tabularx}{0.99\textwidth} { 
   >{\centering\arraybackslash}X 
   >{\centering\arraybackslash}X 
   >{\centering\arraybackslash}X
   >{\centering\arraybackslash}X}
 \toprule \toprule
  & Single-AS & Region-AS & Tetra-AS \\
 \hline
 Mesh partitioning &  --  & By connected region  & By tetrahedron  \\
 Triangle winding order &  --  & Counter-clockwise & Counter-clockwise \\
 Face culling settings &  --  & Front-face culling & Front-face culling \\
 \# of ray payloads & 14 & 16 & 16 \\
 SBT data in addition to surface normals & Tissue labels adjacent to the front and back faces & Both tissue labels and AS handles of adjacent regions & Both tissue labels and AS handles of adjacent regions\\
 \bottomrule \bottomrule
\end{tabularx}
\end{center}
\caption{\label{rtmmc_methods_summary_table} Comparisons between the three AS construction strategies used in the proposed RT-MMC algorithm.}
\end{table}

\section{Results and Discussions}
\subsection{Benchmark configurations}
In this section, we first validate the proposed RT-MMC algorithm, implemented using NVIDIA OptiX 7.5~\cite{OptiXGuide}, comparing its solutions with those of our GPU-accelerated OpenCL-MMC~\cite{Fang2019} using benchmarks from three heterogeneous domains. In benchmark B1 (``cubesph''), a spherical inclusion with a radius of 25~mm and optical properties of absorption coefficient $\mu_a = 0.01~\textrm{mm}^{-1}$, scattering coefficient $\mu_s = 10~\textrm{mm}^{-1}$, anisotropy $g = 0$, and refractive index $n = 1$ is embedded within a $60\times60\times60~\textrm{mm}^3$ cube filled with a material with $\mu_a = 0.005~\textrm{mm}^{-1}, \mu_s = 1~\textrm{mm}^{-1}, g = 0, n = 1.37$. Benchmark B2 (``sphshells'') is detailed in Section~3 of Ref.~\cite{Yan2019dmmc}. Briefly, it consists of three concentric spheres embedded in a $60\times60\times60~\textrm{mm}^3$ cubic domain, with the spheres centered at $[30, 30, 30]$~mm and radii of 10~mm, 23~mm, and 25~mm, respectively. The innermost sphere is filled with a non-scattering medium, with $\mu_a = 0.05~\textrm{mm}^{-1}, \mu_s = 0.0~\textrm{mm}^{-1}, g = 1, n = 1$. The layer between the $10~\textrm{mm}$ and 23~mm spheres mimics gray matter, with optical properties of $\mu_a = 0.02~\textrm{mm}^{-1}, \mu_s = 9.0~\textrm{mm}^{-1}, g = 0.89, n = 1.37$. The layer between the $23~\textrm{mm}$ and $25~\textrm{mm}$ spheres mimics cerebrospinal fluid (CSF), with $\mu_a = 0.004~\textrm{mm}^{-1}, \mu_s = 0.009~\textrm{mm}^{-1}, g = 0.89, n = 1.37$. The region between the $25~\textrm{mm}$ sphere and the cube surface represents the scalp, with $\mu_a = 0.019~\textrm{mm}^{-1}, \mu_s = 7.8~\textrm{mm}^{-1}, g = 0.89, n = 1.37$. In both B1 and B2, the tetrahedral mesh for OpenCL-MMC and Tetra-AS RT-MMC are generated using Tetgen~\cite{Si2015} with the ``-Yq'' flag, achieving a relatively coarse mesh while maintaining good quality~\cite{Yan2019dmmc}. Benchmark B3 (``brain19.5'') uses a brain mesh derived from an MRI brain atlas, with further details provided in Section 3.4 of Ref~\cite{TranYan2020}. All benchmarks simulate $10^8$ photons and are tested on four NVIDIA GPUs representing diverse architectures, including GTX 1080Ti (Pascal), RTX 2080 (Turing), RTX 3090 (Ampere) and RTX 4090 (Ada Lovelace).

\begin{figure}[t]
\centering
    \includegraphics[width=0.6\textwidth]{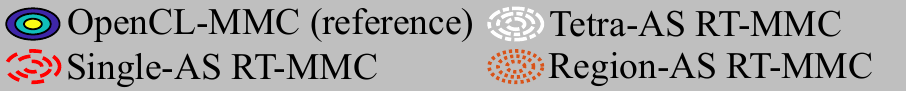}\\[1ex]
    \subfigure[]{
        \includegraphics[width=0.4\textwidth]{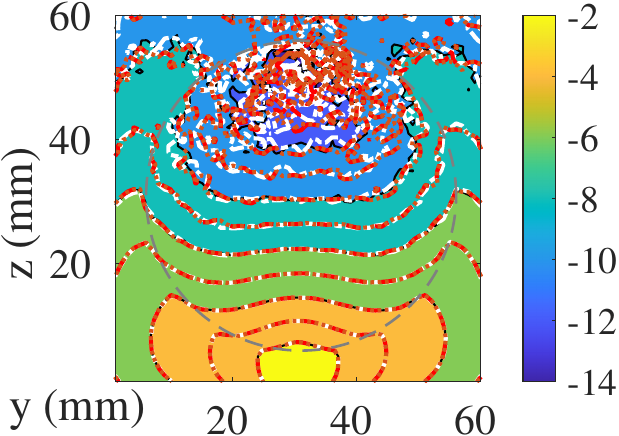}
        \label{rtmmc_results_validation_fig_a}
    }
    \subfigure[]{
        \includegraphics[width=0.4\textwidth]{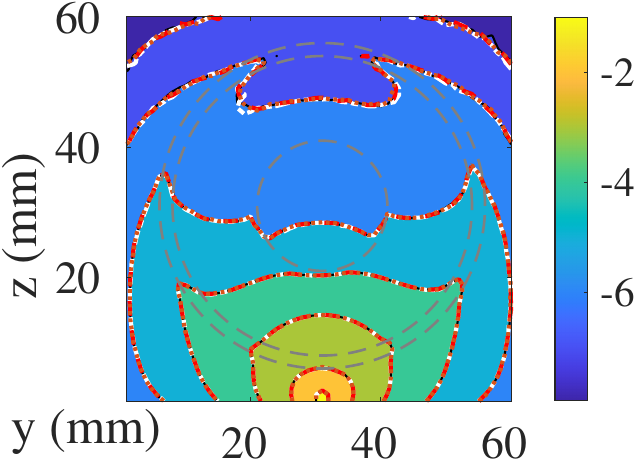}
        \label{rtmmc_results_validation_fig_b}
    }\\[1ex]
    \subfigure[]{
        \includegraphics[width=0.7\textwidth]{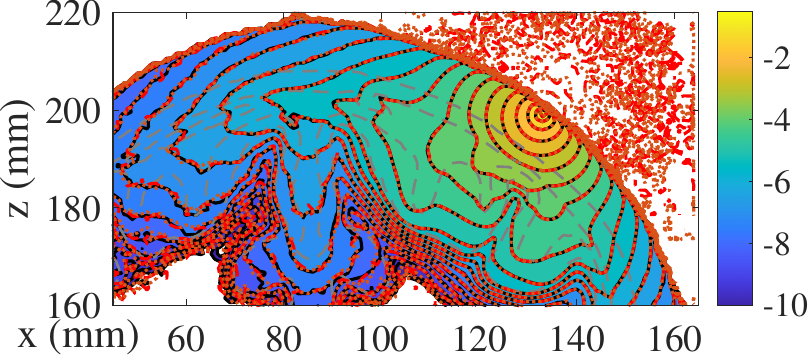}
        \label{rtmmc_results_validation_fig_c}
    }
\caption{\label{rtmmc_results_validation_fig}Fluence ($\text{mm}^{-2}$, in log-10 scale) contour plots for tetrahedral OpenCL-MMC (reference) and three RT-MMC algorithms across various benchmarks: (a) B1, (b) B2, and (c) B3. Gray dashed lines indicate the boundaries of different media.}
\end{figure}

\subsection{Validation of fluence distribution output} \label{rtmmc_results_fluence}
In Fig.~\ref{rtmmc_results_validation_fig}, we compare the spatially-resolved light fluence maps generated by the proposed RT-MMC algorithms against the reference solutions from OpenCL-MMC~\cite{Fang2019}. Overall, the results of all three RT-MMC AS configurations align closely with those of OpenCL-MMC, demonstrating excellent agreement. However, in benchmark B3 [Fig.~\ref{rtmmc_results_validation_fig_c}], the fluence shows unexpected energy deposition outside of the head model, which is assumed to be lossless. This suggests that a small amount of photon packets are not properly terminated when crossing the tissue-air boundary (out-most surface). The underlying cause of this issue is further investigated in Section~\ref{rtmmc_results_future} with potential mitigation strategies.

\subsection{Comparison of simulation speed} \label{rtmmc_results_speedups}
\begin{figure}[h]
\centering
    \includegraphics[width=0.7\textwidth]{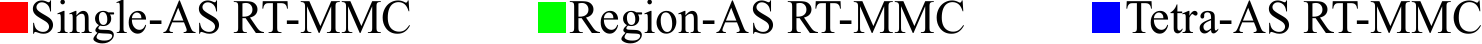}\\[1ex]
    \subfigure[]{
        \includegraphics[width=0.31\textwidth]{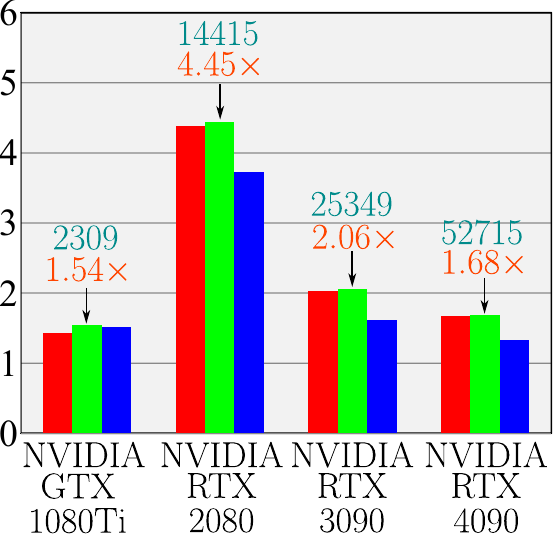}
        \label{rtmmc_results_speed_fig_a}
    }
    \subfigure[]{
        \includegraphics[width=0.31\textwidth]{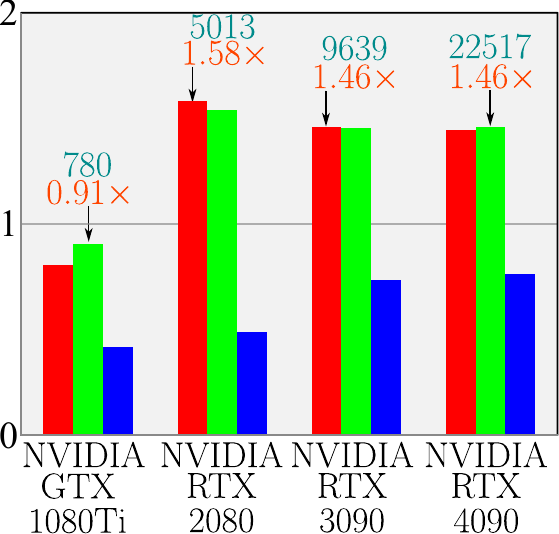}
        \label{rtmmc_results_speed_fig_b}
    }
    \subfigure[]{
        \includegraphics[width=0.31\textwidth]{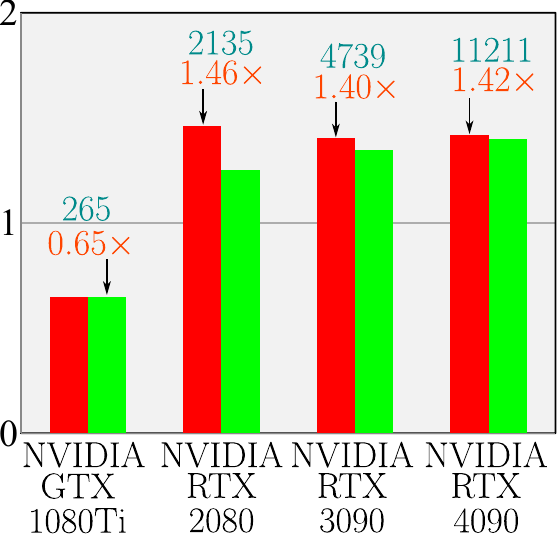}
        \label{rtmmc_results_speed_fig_c}
    }
\caption{\label{rtmmc_results_speed_fig}Comparisons of speedup of RT-MMC algorithms compared to OpenCL-MMC (reference) across four NVIDIA GPUs in benchmarks (a) B1, (b) B2, and (c) B3. The red, green, and blue bars represent the RT-MMC algorithms using Single-AS, Region-AS, and Tetra-AS strategies, respectively, with the highest speedups indicated by black arrows. Blue-colored numbers report the simulation speed in photon/ms, while the orange-colored numbers show the relative speedup over OpenCL-MMC.}
\end{figure}

In Fig.~\ref{rtmmc_results_speed_fig}, we compare the simulation speedups of the RT-MMC algorithms against OpenCL-MMC across three benchmarks and four generations of GPUs. In benchmark B1 [Fig.~\ref{rtmmc_results_speed_fig_a}], Region-AS RT-MMC achieves the highest speedups across all tested GPUs, ranging from $1.54\times$ to $4.45\times$. The Single-AS RT-MMC shows slightly lower but comparable performance. In particular, even on the NVIDIA GTX 1080Ti, which lacks RT cores, RT-MMC outperforms OpenCL-MMC, suggesting that for this relatively simple mesh, the BVH structure is more efficient than the adjacency list used in OpenCL-MMC for neighbor searches within tetrahedra~\cite{Fang2019}. Interestingly, the lowest speedups for the hardware-accelerated RT-MMC algorithms occur on the RTX 4090, despite having the highest number of RT cores.

\begin{table}[h]
\centering
\begin{tabular}{|P{0.9in}|c|c|c|c|c|c|c|c|c|c|c|c|}
\hline
\multirow{3}{*}{\makecell{RT-MMC}} & 
    \multicolumn{12}{c|}{Runtime (ms) for building AS} \\
    \cline{2-13}
    & \multicolumn{3}{c|}{GTX 1080Ti} & \multicolumn{3}{c|}{RTX 2080} & \multicolumn{3}{c|}{RTX 3090} & \multicolumn{3}{c|}{RTX 4090} \\
    \cline{2-13}
    & B1 & B2 & B3 & B1 & B2 & B3 & B1 & B2 & B3 & B1 & B2 & B3 \\
\hline\hline
Single-AS& 2& 2& 11& 1& 1& 2& 1& 1& 2& 1& 1& 1\\
\hline
Region-AS& 3& 5& 700& 1& 2& 110& 1& 2& 81& 1& 2& 99\\
\hline
Tetra-AS& 1149& 9620&  -- & 566& 3888&  -- & 500& 3384&  -- & 462& 3205&  -- \\
\hline
\end{tabular}
\caption{\label{rtmmc_results_asruntime_table}
Summary of the runtime (in ms) for building the AS in benchmarks B1, B2, and B3 across four GPUs.}
\end{table}

In benchmarks B2 [Fig.~\ref{rtmmc_results_speed_fig_b}] and B3 [Fig.~\ref{rtmmc_results_speed_fig_c}], which involve more complex meshes than B1, RT-MMC algorithms are slower than OpenCL-MMC on GTX 1080Ti, due to the lack of RT cores. In comparison, on all GPUs with RT cores present, Single-AS and Region-AS RT-MMC algorithms achieve similar speedups, ranging from $1.46\times$ to $1.58\times$ for B2 and $1.40\times$ to $1.46\times$ for B3. However, Tetra-AS RT-MMC is notably slower than OpenCL-MMC in B2 across all GPUs. For B3, Tetra-AS RT-MMC fails to construct the AS due to extensive memory needs. Additionally, Single-AS and Region-AS RT-MMC outperform Tetra-AS RT-MMC in both processing time and memory usage during AS construction, as detailed in Table~\ref{rtmmc_results_asruntime_table}.

\begin{figure}[!htbp]
\centering
    \subfigure[]{
        \includegraphics[width=0.4\textwidth]{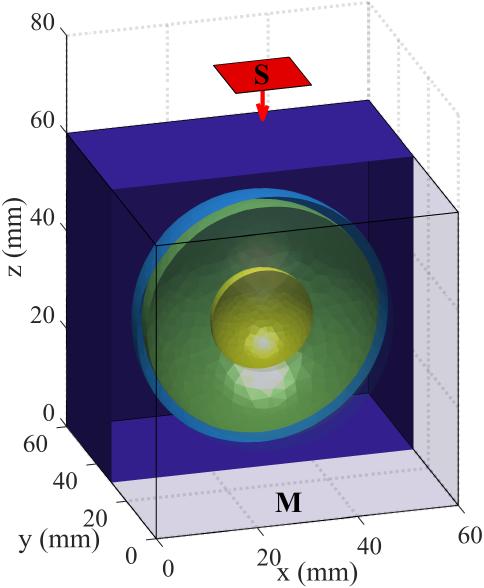}
        \label{rtmmc_results_widefield_fig_a}
    }
    \subfigure[]{
        \includegraphics[width=0.4\textwidth]{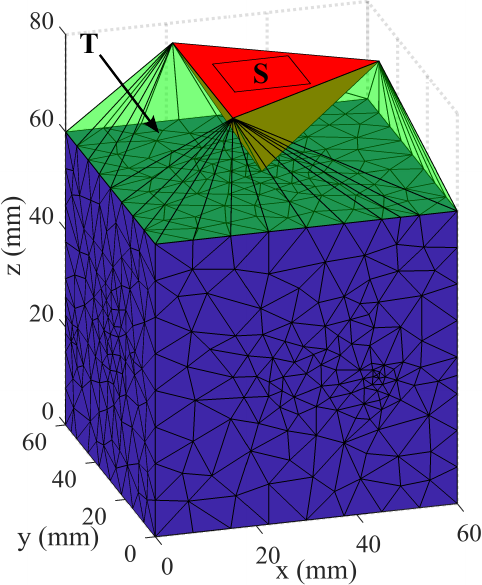}
        \label{rtmmc_results_widefield_fig_b}
    }\\[1ex]
    \subfigure[]{
        \includegraphics[width=0.4\textwidth]{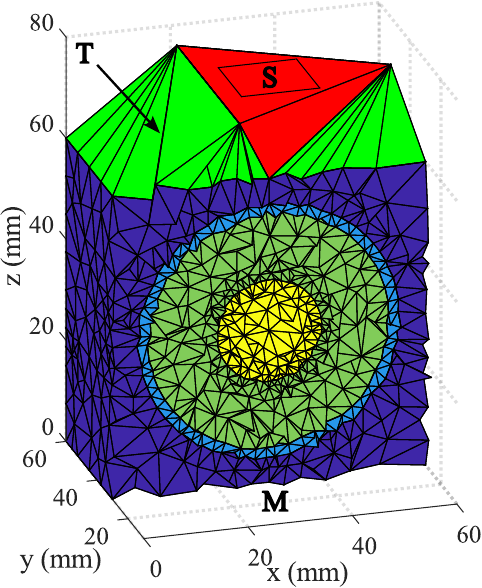}
        \label{rtmmc_results_widefield_fig_c}
    }
    \subfigure[]{
        \includegraphics[width=0.4\textwidth]{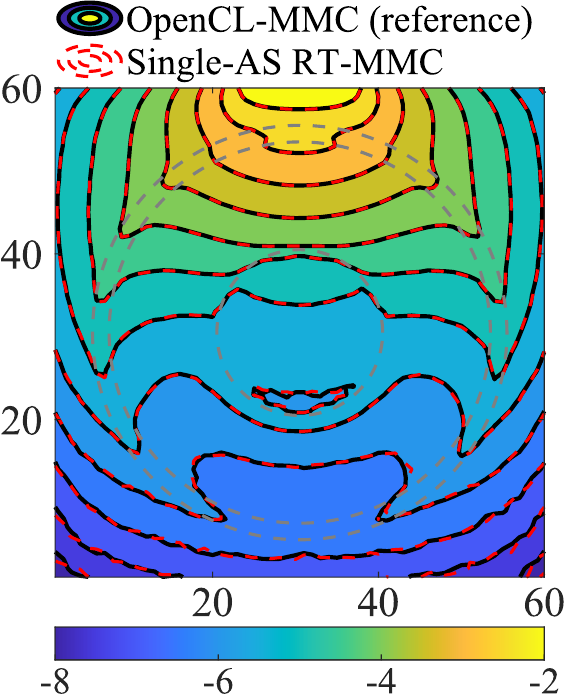}
        \label{rtmmc_results_widefield_fig_d}
    }
\caption{\label{rtmmc_results_widefield_fig}Illustration of benchmark B4, showing
(a) the cross-sectional view of the RT-MMC triangular surface based domain, consisting of the multi-layered tissue mesh domain (``M''), the planar source (``S''), highlighted in red, and the photon launch direction indicated by the red arrow,
(b) extended mesh (``T'') generated by retessellation~\cite{Yao2016} for traditional wide-field MMC simulations, with the source-enclosing element highlighted in red, 
(c) cross-sectional view of the tetrahedral mesh domain for conventional wide-field MMC, and
(d) comparison of fluence contour maps ($\text{mm}^{-2}$, in log-10 scale) for OpenCL-MMC (black) and the Single-AS RT-MMC (red dashed lines). Gray dashed lines indicate media boundaries.}
\end{figure}

\subsection{Demonstration of wide-field source simulations} \label{rtmmc_results_widefield}

To demonstrate the capability of the modified Single-AS RT-MMC in simulating wide-field sources, we create a new validation benchmark B4 by replacing the pencil beam source by a ``planar'' source in benchmark B2 (``sphereshells''). The planar source uniformly launches photons from a $15\times15~\textrm{mm}^2$ area centered at $[30, 30, 80]~\textrm{mm}$ with a launch direction pointing toward the -$z$-axis [see Fig.~\ref{rtmmc_results_widefield_fig_a}]. We evaluated this simulation configuration using both OpenCL-MMC and the modified Single-AS RT-MMC on a desktop equipped with an NVIDIA RTX 2080 Super GPU, simulating $10^8$ photons in each run. Figures~\ref{rtmmc_results_widefield_fig_b}-\ref{rtmmc_results_widefield_fig_c} depict the extended mesh domain produced by the mesh retessellation required for traditional MMC~\cite{Yao2016} -- an overhead that is completely eliminated in the modified Single-AS RT-MMC. As shown in Fig.~\ref{rtmmc_results_widefield_fig_d}, the fluence maps generated by the two solvers exhibit excellent agreement. Moreover, the modified single-AS RT-MMC achieves a speedup of $1.53\times$ over traditional MMC (5,079 photons/ms vs. 3,322 photons/ms), consistent with the trends reported in Fig.~\ref{rtmmc_results_speed_fig_b}, confirming that the overhead of dynamically determining the initial medium type is negligible.

\subsection{Investigation of photon leakage across tissue boundary} \label{rtmmc_results_future}
We further investigated the photon leakage issue observed in Fig.~\ref{rtmmc_results_validation_fig_c} in Section~\ref{rtmmc_results_fluence} and identified its origin as a numerical precision related error in the GPU's built-in intersection shader~\cite{photonleakage}. Specifically, the single-precision arithmetic used for ray-triangle intersection calculations can introduce rounding errors, causing small positive $d_{min}$ (i.e., the distance to the closest intersection) to be rounded to zero or even negative values. When $d_{min}$ becomes less than or equal to 0, the built-in intersection shader failed to detect the intersection and falsely trigger the ``miss'' event. In such cases, a photon may cross a boundary undetected, propagating into a new tissue region while erroneously retaining the optical properties of the previous region. Unfortunately, to our knowledge, OptiX does not provide a built-in mechanism to detect such precision-induced errors~\cite{OptiXGuide}.

\begin{figure}[t]
\centering
    \includegraphics[width=0.6\textwidth]{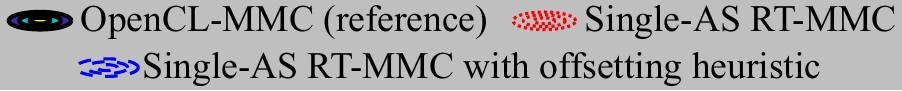}\\[1ex]
    \subfigure[]{
        \includegraphics[width=0.45\textwidth]{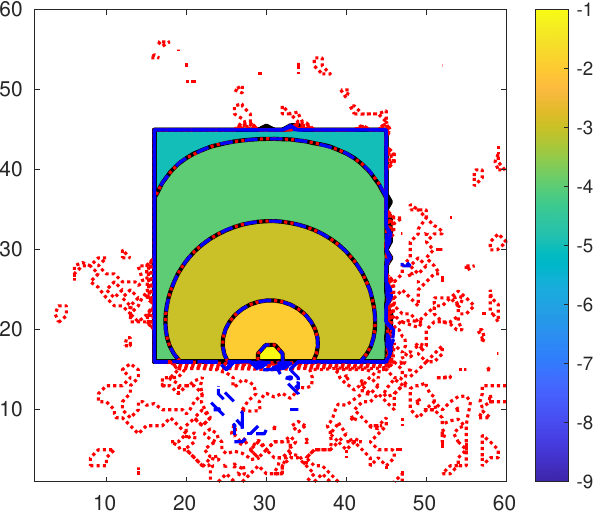}
        \label{rtmmc_results_heuristic_fig_a}
    }
    \subfigure[]{
        \includegraphics[width=0.45\textwidth]{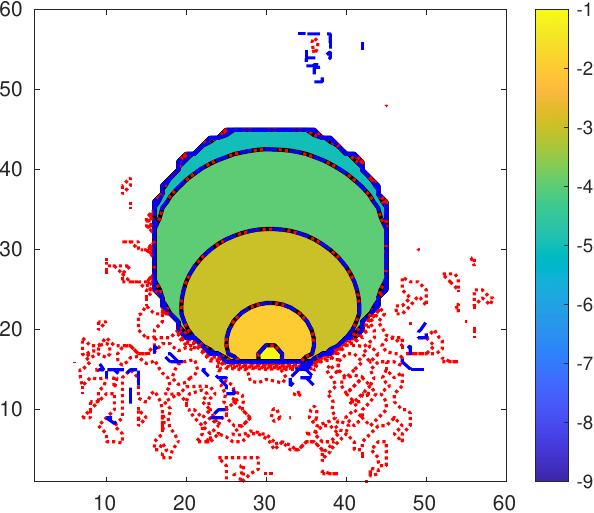}
        \label{rtmmc_results_heuristic_fig_b}
    }
\caption{\label{rtmmc_results_heuristic_fig}Fluence ($\text{mm}^{-2}$, in log-10 scale) contour plots comparing OpenCL-MMC (black solid), Single-AS RT-MMC (red dotted), and Single-AS RT-MMC with the retraction heuristic (blue dashed) in benchmarks: (a) B5 and (b) B6.}
\end{figure}

To quantify and characterize the severity of this issue, we designed two additional benchmarks, B5 and B6 (see Fig.~\ref{rtmmc_results_heuristic_fig}). Both benchmarks feature a cubic domain of size $60\times60\times60~\textrm{mm}^3$, filled with air ($\mu_a = 0.0~\textrm{mm}^{-1}$, $\mu_s = 0.0~\textrm{mm}^{-1}$, $g = 1.0$, $n = 1.0$). In B5, a $30\times30\times30~\textrm{mm}^3$ cubic inclusion filled with scattering material ($\mu_a = 0.005~\textrm{mm}^{-1}$, $\mu_s = 10.0~\textrm{mm}^{-1}$, $g = 0.9$, $n = 1.37$) is centered at [30, 30, 30]~mm. In B6, a spherical inclusion with a radius of 15~mm, filled with the same scattering material, is also centered at [30, 30, 30]~mm. In both benchmarks, a pencil beam source is positioned at [30, 30, 15]~mm and directed along the +$z$-axis. The inclusion sizes and positions were carefully selected to ensure that their boundaries could be described parametrically.

To capture those ``leaked'' photons, we modified the miss shader in the Single-AS RT-MMC to detect photons that have exited into the ambient air while retaining the optical properties of the inclusion. The trajectories of these leaked photons were tracked and analyzed. In this experiment, $10^8$ photons were simulated for both benchmarks, with B5 reporting 1,541 leaked photons ($0.001541\%$) and B6 reporting 591 leaked photons ($0.000591\%$). Analysis of the trajectories showed that leakage always occurred after a scattering event, where the photon's direction change caused the next $d_{min}$ value to reach the precision limits of single-precision arithmetic.

\subsubsection{Potential mitigation strategy using a retraction heuristic}
Based on these findings, we propose a heuristic to mitigate the photon leakage issue. For each ray-tracing call, we introduce a tiny retraction length $\Delta$ to the ray's origin along its direction. The modified origin is calculated as $\mathbf{p}' = \mathbf{p} - \Delta \cdot \mathbf{v}$, where $\mathbf{p}$ is the unmodified origin of the ray, $\mathbf{v}$ is the direction of the ray, and $\Delta$ is the retraction distance. This adjustment effectively increases $d_{min}$ by $\Delta$, reducing the likelihood of precision-limit errors. To account for the retraction distance, $d_{max}$ is adjusted to $s/{\mu_s}+\Delta$. Additionally, in the closest-hit shader, $d_{min}$ is decreased by $\Delta$ to restore the actual travel distance.

We evaluated the proposed heuristic, using $\Delta = 10^{-5}$~mm, by simulating $10^8$ photons in benchmarks B5 and B6 with the Single-AS RT-MMC. The results showed a substantial reduction in photon leakage, with leaked photons decreasing from 1,541 to 49 in B5 and from 591 to 21 in B6 -- a nearly 30-fold reduction. This heuristic shows no impact on overall performance. However, leakage was not completely eliminated, as demonstrated by the non-zero light fluence outside the inclusion regions in Fig.~\ref{rtmmc_results_heuristic_fig}. This suggests that while the heuristic significantly mitigates the issue, it may not be fully robust, and further investigation is needed.

\section{Conclusion}
In summary, we present RT-MMC, an RT-core accelerated mesh-based Monte Carlo photon simulation algorithm designed to leverage modern GPU ray-tracing frameworks. The proposed algorithm encapsulates key steps in physically accurate 3D photon transport simulations within standardized RT pipelines and can be readily generalized to many contemporary ray-tracing platforms including NVIDIA OptiX (current implementation), Microsoft DXR~\cite{DXRGuide,VulkanRayTracingSIGGRAPH}, Apple Metal~\cite{MetalRayTracing}, and Vulkan~\cite{VulkanGuide}. We tested three different strategies (Single-AS, Region-AS and Tetra-AS) for creating AS based on domain mesh models and compared their runtimes across various NVIDIA GPU architectures. The proposed RT-MMC algorithm and the OptiX-based software were systematically evaluated using diverse benchmarks ranging from simple to highly complex heterogeneous tissue domains. The output fluence maps showed excellent agreement with traditional MMC methods while achieving significant speedups ranging from $1.46\times$ to $4.45\times$ across various GPU architectures. Comparing across the three AS configurations, the Single-AS approach requires the least amount of GPU resources while achieving the highest simulation speed in many of the benchmarks. Region-AS shows similar performance with a slight increase in AS construction runtime.

Moreover, RT-MMC greatly simplifies the simulation of wide-field and volumetric sources, which are traditionally challenging for conventional MMC. With minimal modifications, we generalize the Singe-AS RT-MMC algorithm to support these source types with negligible overhead, achieving excellent agreement with conventional MMC while eliminating mesh preprocessing. However, minor discrepancies were observed in some benchmarks due to a small fraction ($<0.002\%$) of ``leaked'' photons -- instances where photons are not properly terminated at tissue boundaries. Further investigation attributed this issue to the precision limits of the OptiX built-in ray-triangle intersection shader. A simple heuristic was explored, resulting in a 30-fold reduction in the number of leaked photons at tissue boundaries. Future work will focus on addressing these limitations and expanding towards other ray-tracing frameworks.

\section*{Disclosures}
No conflicts of interest, financial or otherwise, are declared by the authors.

\section* {Code, Data, and Materials Availability}
The RT-MMC algorithm has been integrated into our widely used open-source MMC simulator and is freely available at \url{https://mcx.space/}. The source code of the software can be accessed at \url{https://github.com/fangq/mmc/}.

\acknowledgments
This research is supported by the National Institutes of Health (NIH) grants R01-GM114365, R01-CA204443 and R01-EB026998. We would like to thank David Hart from NVIDIA for helping with the investigation of the photon leakage issue discussed in Section~\ref{rtmmc_results_future}. 

\bibliography{Bibliography}   

@misc{photonleakage,
    author = {{David Hart}},
    title = {{An example illustrating single precision limits in OptiX ray-triangle intersection}},
    institution = {NVIDIA},
    year = 2022,
    howpublished = {\url{https://forums.developer.nvidia.com/t/two-questions-1-payloadtype-semantics-2-ray-triangle-intersection/227569/10}},
    note = {Accessed: 2024}
}

@article{Doronin2011,
    author = "A. Doronin and I. Meglinski",
    title = {{Online object oriented Monte Carlo computational tool for the needs of biomedical optics}},
    journal= "Biomed. Opt. Express",
    volume = "2",
    number = "9",
    pages = "2461-2469",
    year = "2011",
}

@article{Jonsson2020,
    author = {Joakim J\"{o}nsson and Edouard Berrocal},
    journal = {Opt. Express},
    number = {25},
    pages = {37612--37638},
    publisher = {Optica Publishing Group},
    title = {{Multi-Scattering software: part I: online accelerated Monte Carlo simulation of light transport through scattering media}},
    volume = {28},
    month = {Dec},
    year = {2020},
    url = {https://opg.optica.org/oe/abstract.cfm?URI=oe-28-25-37612},
    doi = {10.1364/OE.404005},
}

@INPROCEEDINGS{Salmon2019,
    author={Salmon, Justin and McIntosh-Smith, Simon},
    booktitle={2019 IEEE/ACM Performance Modeling, Benchmarking and Simulation of High Performance Computer Systems (PMBS)}, 
    title={Exploiting Hardware-Accelerated Ray Tracing for Monte Carlo Particle Transport with OpenMC}, 
    year={2019},
    volume={},
    number={},
    pages={19-29},
    doi={10.1109/PMBS49563.2019.00008}
}

@misc{Kilgariff2018,
    author = {Emmett Kilgariff and Henry Moreton and Nick Stam and Brandon Bell},
    title = {{NVIDIA Turing Architecture In-Depth}},
    year = 2018,
    url = {https://developer.nvidia.com/blog/nvidia-turing-architecture-in-depth},
    urldate = {2024-09-01}
}

@article{McMillan2022,
    author = {Lewis McMillan and Graham D. Bruce and Kishan Dholakia},
    title = {{Meshless Monte Carlo radiation transfer method for curved geometries using signed distance functions}},
    volume = {27},
    journal = {Journal of Biomedical Optics},
    number = {8},
    publisher = {SPIE},
    pages = {083003},
    year = {2022},
    doi = {10.1117/1.JBO.27.8.083003},
    URL = {https://doi.org/10.1117/1.JBO.27.8.083003}
}

@misc{mcxyz,
    author = {Steven Jacques and Ting Li and Scott Prahl},
    title = {{mcxyz.c, a 3D Monte Carlo simulation of heterogeneous tissues}},
    year = 2019,
    url = {omlc.org/software/mc/mcxyz},
    urldate = {2024-09-01}
}

@article{Marti2021,
    author = {Dominik Marti and Rikke N. N. Aasbjerg and Peter E. E. Andersen and Anders K. K. Hansen},
    title = {{MCmatlab: an open-source, user-friendly, MATLAB-integrated three-dimensional Monte Carlo light transport solver with heat diffusion and tissue damage (Erratum)}},
    volume = {26},
    journal = {Journal of Biomedical Optics},
    number = {1},
    publisher = {SPIE},
    pages = {019804},
    year = {2021},
    doi = {10.1117/1.JBO.26.1.019804},
    URL = {https://doi.org/10.1117/1.JBO.26.1.019804}
}

@article{Hielscher1998,
    doi = {10.1088/0031-9155/43/5/017},
    url = {https://dx.doi.org/10.1088/0031-9155/43/5/017},
    year = {1998},
    month = {may},
    publisher = {},
    volume = {43},
    number = {5},
    pages = {1285},
    author = {Andreas H Hielscher and  Raymond E Alcouffe and  Randall L Barbour},
    title = {Comparison of finite-difference transport and diffusion calculations for photon migration in homogeneous and heterogeneous tissues},
    journal = {Physics in Medicine \& Biology},
}

@article{Okada2003,
    author = {Eiji Okada and David T. Delpy},
    journal = {Appl. Opt.},
    number = {16},
    pages = {2906--2914},
    publisher = {Optica Publishing Group},
    title = {Near-infrared light propagation in an adult head model. I. Modeling of low-level scattering in the cerebrospinal fluid layer},
    volume = {42},
    month = {Jun},
    year = {2003},
    url = {https://opg.optica.org/ao/abstract.cfm?URI=ao-42-16-2906},
    doi = {10.1364/AO.42.002906},
}

@article{Custo2006,
    author = {Anna Custo and William M. Wells III and Alex H. Barnett and Elizabeth M. C. Hillman and David A. Boas},
    journal = {Appl. Opt.},
    number = {19},
    pages = {4747--4755},
    publisher = {Optica Publishing Group},
    title = {Effective scattering coefficient of the cerebral spinal fluid in adult head models for diffuse optical imaging},
    volume = {45},
    month = {Jul},
    year = {2006},
    url = {https://opg.optica.org/ao/abstract.cfm?URI=ao-45-19-4747},
    doi = {10.1364/AO.45.004747},
}

@article{Fang2011,
    author = {Qianqian Fang},
    journal = {Biomed. Opt. Express},
    number = {5},
    pages = {1258--1264},
    publisher = {Optica Publishing Group},
    title = {{Comment on ``A study on tetrahedron-based inhomogeneous Monte-Carlo optical simulation''}},
    volume = {2},
    month = {May},
    year = {2011},
    url = {https://opg.optica.org/boe/abstract.cfm?URI=boe-2-5-1258},
    doi = {10.1364/BOE.2.001258},
}

@book{pharr2023physically,
    title={Physically Based Rendering, fourth edition: From Theory to Implementation},
    author={Pharr, M. and Jakob, W. and Humphreys, G.},
    isbn={9780262048026},
    lccn={2022014718},
    url={https://books.google.com/books?id=ENSMEAAAQBAJ},
    year={2023},
    publisher={MIT Press}
}

@inproceedings{Stich2009,
    author = {Stich, Martin and Friedrich, Heiko and Dietrich, Andreas},
    title = {Spatial splits in bounding volume hierarchies},
    year = {2009},
    isbn = {9781605586038},
    publisher = {Association for Computing Machinery},
    address = {New York, NY, USA},
    url = {https://doi.org/10.1145/1572769.1572771},
    doi = {10.1145/1572769.1572771},
    booktitle = {Proceedings of the Conference on High Performance Graphics 2009},
    pages = {7–13},
    numpages = {7},
    location = {New Orleans, Louisiana},
    series = {HPG '09}
}

@book{Marrs2021,
    title = {Ray Tracing Gems II},
    editor = {Adam Marrs and Peter Shirley and Ingo Wald},
    publisher = {Apress},
    year = {2021},
    note ={\url{http://raytracinggems.com/rtg2}},
}

@article{Yan2020,
    author = {Shijie Yan and Qianqian Fang},
    journal = {Biomed. Opt. Express},
    number = {11},
    pages = {6262--6270},
    publisher = {Optica Publishing Group},
    title = {Hybrid mesh and voxel based Monte Carlo algorithm for accurate and efficient photon transport modeling in complex bio-tissues},
    volume = {11},
    month = {Nov},
    year = {2020},
    url = {https://opg.optica.org/boe/abstract.cfm?URI=boe-11-11-6262},
    doi = {10.1364/BOE.409468},
}

@article{Parker2010,
    author = {Parker, Steven G. and Bigler, James and Dietrich, Andreas and Friedrich, Heiko and Hoberock, Jared and Luebke, David and McAllister, David and McGuire, Morgan and Morley, Keith and Robison, Austin and Stich, Martin},
    title = {{OptiX: a general purpose ray tracing engine}},
    year = {2010},
    issue_date = {July 2010},
    publisher = {Association for Computing Machinery},
    address = {New York, NY, USA},
    volume = {29},
    number = {4},
    issn = {0730-0301},
    url = {https://doi.org/10.1145/1778765.1778803},
    doi = {10.1145/1778765.1778803},
    journal = {ACM Trans. Graph.},
    month = {jul},
    articleno = {66},
    numpages = {13}
}

@article{Wang1995,
    author = "L. V. Wang and S. L. Jacques and L. Zheng",
    title = {{MCML-Monte Carlo modeling of light transport in multi-layered tissues}},
    journal = "Comput. Methods Progr. Biomed.",
    volume = "47",
    number = "2",
    pages = "131-146",
    year = "1995",
}

@article{Yao2016,
    author = "R. Yao and X. Intes and Q. Fang",
    title = {{Generalized mesh-based Monte Carlo for wide-field illumination and detection via mesh retessellation}},
    journal = "Biomed. Opt. Express",
    volume = "7",
    number = "1",
    pages = "171-184",
    year = "2016",
}

@article{Fang2009mcx,
    author = "Q. Fang and D. A. Boas",
    title = {{Monte Carlo simulation of photon migration in 3D turbid media accelerated by graphics processing units}},
    journal = "Opt. Express",
    volume = "17",
    number = "22",
    pages = "20178-20190",
    year = "2009",
}

@article{Fang2010,
    author = "Q. Fang",
    title = {{Mesh-based Monte Carlo method using fast ray- tracing in Pl\"ucker coordinates}},
    journal = "Biomed. Opt. Express",
    volume = "1",
    number = "1",
    pages = "165-175",
    year = "2010",
}

@article{Fang2012,
    author = "Q. Fang and D. Kaeli",
    title = {{Accelerating mesh-based Monte Carlo method on modern CPU architectures}},
    journal= "Biomed. Opt. Express",
    volume = "3",
    number = "12",
    pages = "3223-3230",
    year = "2012",
}

@article{Si2015,
    author = "H. Si",
    title = {Tet{G}en, a {D}elaunay-Based Quality Tetrahedral Mesh Generator},
    journal= "AMC Trans. Math. Software",
    volume = "41",
    number = "2",
    year = "2015",
}

@article{Boas2002,
    author = {D. A. Boas and J. P. Culver and J. J. Stott and A. K. Dunn},
    journal = {Opt. Express},
    number = {3},
    pages = {159--170},
    publisher = {OSA},
    title = {{Three dimensional Monte Carlo code for photon migration through complex heterogeneous media including the adult human head}},
    volume = {10},
    month = {Feb},
    year = {2002},
    url = {http://www.opticsexpress.org/abstract.cfm?URI=oe-10-3-159},
    doi = {10.1364/OE.10.000159},
}

@article{TranYan2020,
author = {Anh Phong Tran and Shijie Yan and Qianqian Fang},
title = {{Improving model-based functional near-infrared spectroscopy analysis using mesh-based anatomical and light-transport models}},
volume = {7},
journal = {Neurophotonics},
number = {1},
publisher = {SPIE},
pages = {015008},
year = {2020},
doi = {10.1117/1.NPh.7.1.015008},
URL = {https://doi.org/10.1117/1.NPh.7.1.015008}
}

@article{Shen2010,
    author = "H. Shen and G. Wang",
    title = {{A tetrahedron-based inhomogeneous Monte Carlo optical simulator}},
    journal= "Phys. Med. Biol.",
    volume = "55",
    number = "4",
    pages = "947-962",
    year = "2010",
}

@article{Fang2019,
author = {Qianqian Fang and Shijie Yan},
title = {{Graphics processing unit-accelerated mesh-based Monte Carlo photon transport simulations}},
volume = {24},
journal = {Journal of Biomedical Optics},
number = {11},
publisher = {SPIE},
pages = {115002},
year = {2019},
doi = {10.1117/1.JBO.24.11.115002},
URL = {https://doi.org/10.1117/1.JBO.24.11.115002}
}

@article{Yu2018,
    author = { Leiming Yu and Fanny Nina-Paravecino and David R. Kaeli and Qianqian Fang},
    title = {{Scalable and massively parallel Monte Carlo photon transport simulations for heterogeneous computing platforms}},
    volume = {23},
    journal = {Journal of Biomedical Optics},
    number = {1},
    pages = {1 - 4 - 4},
    year = {2018},
    doi = {10.1117/1.JBO.23.1.010504},
    URL = {https://doi.org/10.1117/1.JBO.23.1.010504},
    eprint = {}
}

@article{Yuan2020,
author = {Yaoshen Yuan and Paolo Cassano and Matthew Pias and Qianqian Fang},
title = {{Transcranial photobiomodulation with near-infrared light from childhood to elderliness: simulation of dosimetry}},
volume = {7},
journal = {Neurophotonics},
number = {1},
publisher = {SPIE},
pages = {015009},
year = {2020},
doi = {10.1117/1.NPh.7.1.015009},
URL = {https://doi.org/10.1117/1.NPh.7.1.015009}
}

@article{Alerstam2008,
author = {Erik Alerstam and Tomas Svensson and Stefan Andersson-Engels},
title = {{Parallel computing with graphics processing units for high-speed Monte Carlo simulation of photon migration}},
volume = {13},
journal = {Journal of Biomedical Optics},
number = {6},
publisher = {SPIE},
pages = {060504},
year = {2008},
doi = {10.1117/1.3041496},
URL = {https://doi.org/10.1117/1.3041496}
}

@article{Ren2010,
    author = {Nunu Ren and Jimin Liang and Xiaochao Qu and Jianfeng Li and Bingjia Lu and Jie Tian},
    journal = {Opt. Express},
    number = {7},
    pages = {6811--6823},
    publisher = {Optica Publishing Group},
    title = {{GPU-based Monte Carlo simulation for light propagation in complex heterogeneous tissues}},
    volume = {18},
    month = {Mar},
    year = {2010},
    url = {https://opg.optica.org/oe/abstract.cfm?URI=oe-18-7-6811},
    doi = {10.1364/OE.18.006811},
}

@article{Binzoni2008,
    title = {{Light transport in tissue by 3D Monte Carlo: Influence of boundary voxelization}},
    journal = {Computer Methods and Programs in Biomedicine},
    volume = {89},
    number = {1},
    pages = {14-23},
    year = {2008},
    issn = {0169-2607},
    doi = {https://doi.org/10.1016/j.cmpb.2007.10.008},
    url = {https://www.sciencedirect.com/science/article/pii/S0169260707002350},
    author = {T. Binzoni and T.S. Leung and R. Giust and D. Rüfenacht and A.H. Gandjbakhche},
}

@article{Majaron2015,
author = {Boris Majaron and Matija Milanič and Jan Premru},
title = {{Monte Carlo simulation of radiation transport in human skin with rigorous treatment of curved tissue boundaries}},
volume = {20},
journal = {Journal of Biomedical Optics},
number = {1},
publisher = {SPIE},
pages = {015002},
year = {2015},
doi = {10.1117/1.JBO.20.1.015002},
URL = {https://doi.org/10.1117/1.JBO.20.1.015002}
}

@article {Ding2016,
    Title = {Influence of surface curvature on light-based nondestructive measurement of stone fruit},
    Author = {Ding, Chizhu and Jianjun Chen and Shuning Shi and Wei Wei and Zuojun Tan},
    DOI = {10.1016/j.compag.2015.12.008},
    Volume = {121},
    Month = {February},
    Year = {2016},
    Journal = {Computers and electronics in agriculture},
    ISSN = {0168-1699},
    Pages = {200-206},
    URL = {https://doi.org/10.1016/j.compag.2015.12.008},
}

@article{Periyasamy2014,
author = {Vijitha Periyasamy and Manojit Pramanik},
title = {{Monte Carlo simulation of light transport in turbid medium with embedded object—spherical, cylindrical, ellipsoidal, or cuboidal objects embedded within multilayered tissues}},
volume = {19},
journal = {Journal of Biomedical Optics},
number = {4},
publisher = {SPIE},
pages = {045003},
year = {2014},
doi = {10.1117/1.JBO.19.4.045003},
URL = {https://doi.org/10.1117/1.JBO.19.4.045003}
}

@article{Zhang2015,
    author = {Yong Zhang and Bin Chen and Dong Li and Guo-Xiang Wang},
    title = {Efficient and Accurate Simulation of Light Propagation in Bio-Tissues Using the Three-Dimensional Geometric Monte Carlo Method},
    journal = {Numerical Heat Transfer, Part A: Applications},
    volume = {68},
    number = {8},
    pages = {827-846},
    year  = {2015},
    publisher = {Taylor & Francis},
    doi = {10.1080/10407782.2015.1023140},
    URL = {https://doi.org/10.1080/10407782.2015.1023140},
    eprint = {https://doi.org/10.1080/10407782.2015.1023140}
}

@article{Margallo2007,
    author = "E. Margallo-Balb\'as and P. J. French",
    title = {{Shape based Monte Carlo code for light transport in complex heterogeneous tissues}},
    journal = "Opt. Express",
    volume = "15",
    number = "21",
    pages = "14086-14098",
    year = "2007",
}

@article{Young-Schultz2019,
    author = {Tanner Young-Schultz and Stephen Brown and Lothar Lilge and Vaughn Betz},
    journal = {Biomed. Opt. Express},
    number = {9},
    pages = {4711--4726},
    publisher = {OSA},
    title = {{FullMonteCUDA: a fast, flexible, and accurate GPU-accelerated Monte Carlo simulator for light propagation in turbid media}},
    volume = {10},
    month = {Sep},
    year = {2019},
    url = {http://opg.optica.org/boe/abstract.cfm?URI=boe-10-9-4711},
    doi = {10.1364/BOE.10.004711},
}

@article{Tran2020,
author = {Anh Phong Tran and Steven L. Jacques},
title = {{Modeling voxel-based Monte Carlo light transport with curved and oblique boundary surfaces}},
volume = {25},
journal = {Journal of Biomedical Optics},
number = {2},
publisher = {SPIE},
pages = {025001},
year = {2020},
doi = {10.1117/1.JBO.25.2.025001},
URL = {https://doi.org/10.1117/1.JBO.25.2.025001}
}

@manual{DXRGuide,
    title = {{DirectX Raytracing (DXR) Functional Spec}},
    year = {2023},
    month = {Jan},
    organization = {Microsoft Corp.},
    note = {Version 1.20}
}

@manual{OptiXGuide,
    title = {{NVIDIA OptiX 7.5 Programming Guide}},
    year = {2022},
    month = {May},
    organization = {NVIDIA Corp.},
    note = {Version 1.13}
}

@article{Yan2019dmmc,
  author = {Yan, Shijie and Tran, Anh Phong and Fang, Qianqian},
  title = {{A dual-grid mesh-based Monte Carlo algorithm for efficient photon transport simulations in complex 3-D media}},
  journal = {Journal of Biomedical Optics},
  volume = {24},
  number = {2},
  pages = {020503},
  year = {2019},
  month = {february},
  doi = {10.1117/1.JBO.24.2.020503},
  url = {https://doi.org/10.1117/1.JBO.24.2.020503},
}

@article{Alerstam2010,
	author = "E. Alerstam and W. C. Y. Lo and T. D. Han and J. Rose and S. Andersson-Engels and L. Lilge",
	title = {{Next-generation acceleration and code optimization for light transport in turbid media using GPUs}},
	journal = "Biomed. Opt. Express",
	volume = "1",
	number = "2",
	pages = "658-675",
	year = "2010",
}

@misc{MetalRayTracing,
  author = {{Apple Inc.}},
  title = {Metal Ray Tracing},
  year = {2022},
  howpublished = {\url{https://developer.apple.com/documentation/metalperformanceshaders/metal_for_accelerating_ray_tracing}},
  note = {Accessed: 2024}
}

@inproceedings{VulkanRayTracingSIGGRAPH,
  author = {Wyman, Chris},
  title = {Introduction to {DirectX} Raytracing and {Vulkan} Ray Tracing},
  booktitle = {ACM SIGGRAPH 2019 Courses},
  year = {2019},
  publisher = {ACM},
  doi = {10.1145/3305366.3328083}
}

@misc{VulkanGuide,
  author = {{Khronos Group}},
  title = {Vulkan Ray Tracing Final Specification Release},
  year = {2020},
  howpublished = {\url{https://www.khronos.org/blog/vulkan-ray-tracing-final-specification-release}},
  note = {Accessed: 2024}
}

@inproceedings{jensen2001practical,
  author = {Jensen, Henrik Wann and Marschner, Stephen R. and Levoy, Marc and Hanrahan, Pat},
  title = {A Practical Model for Subsurface Light Transport},
  booktitle = {Proceedings of the 28th Annual Conference on Computer Graphics and Interactive Techniques},
  series = {SIGGRAPH '01},
  year = {2001},
  pages = {511--518},
  publisher = {ACM},
  address = {New York, NY, USA},
  doi = {10.1145/383259.383319}
}

@article{bentley1975multidimensional,
  author = {Bentley, Jon Louis},
  title = {Multidimensional binary search trees used for associative searching},
  journal = {Communications of the ACM},
  year = {1975},
  volume = {18},
  number = {9},
  pages = {509--517},
  doi = {10.1145/361002.361007}
}

@article{meagher1982geometric,
  author = {Meagher, Donald},
  title = {Geometric modeling using octree encoding},
  journal = {Computer Graphics and Image Processing},
  year = {1982},
  volume = {19},
  number = {2},
  pages = {129--147},
  doi = {10.1016/0146-664X(82)90104-6}
}

@article{kay1986ray,
  author = {Kay, Timothy L. and Kajiya, James T.},
  title = {Ray tracing complex scenes},
  journal = {ACM SIGGRAPH Computer Graphics},
  year = {1986},
  volume = {20},
  number = {4},
  pages = {269--278},
  doi = {10.1145/15886.15916}
}

@book{botsch2010polygon,
  author = {Botsch, Mario and Kobbelt, Leif and Pauly, Mark and Alliez, Pierre and Lévy, Bruno},
  title = {Polygon Mesh Processing},
  year = {2010},
  publisher = {A K Peters/CRC Press},
  isbn = {978-1568814261}
}

@article{Vigna2017,
title = {Further scramblings of Marsaglia’s xorshift generators},
journal = {Journal of Computational and Applied Mathematics},
volume = {315},
pages = {175-181},
year = {2017},
issn = {0377-0427},
doi = {https://doi.org/10.1016/j.cam.2016.11.006},
url = {https://www.sciencedirect.com/science/article/pii/S0377042716305301},
author = {Sebastiano Vigna}
}
\bibliographystyle{spiejour}   

\vspace{2ex}\noindent\textbf{Shijie Yan} is a PhD candidate in electrical engineering at Northeastern University. He received his BE degree in information science and engineering from Southeast University, China, in 2013 and MS degree in electrical and computer engineering from Northeastern University in 2017. His research interests include Monte Carlo photon transport simulation algorithms, parallel computing, GPU programming and optimization.

\vspace{2ex}\noindent\textbf{Qianqian Fang}, PhD, is an associate professor in the Bioengineering Department at Northeastern University, Boston, Massachusetts, United States. He received his PhD from the Thayer School of Engineering, Dartmouth College, in 2005. He then joined Massachusetts General Hospital and became an instructor of radiology in 2009 and an assistant professor of radiology in 2012, before joining Northeastern University in 2015 as an assistant professor. His research interests include translational medical imaging devices, multi-modal imaging, image reconstruction algorithms, scientific data sharing, and high-performance computing tools to facilitate the development of next-generation imaging platforms..

\end{spacing}
\end{sloppypar}
\end{document}